\documentclass[preprint,proceedings]{rmaa}

\usepackage{paralist}
\usepackage{rmaacite}

\usepackage{psfrag,color}

\def\ha{H$\alpha$}

\SetYear{2002}
\SetConfTitle{Galaxy Evolution: Theory and Observations}

\title{Galaxy Evolution: Internally or Externally Driven?} 

\author{
  M. L. Balogh\altaffilmark{1} 
  and R.G. Bower\altaffilmark{1}
}
\altaffiltext{1}{Department of Physics, University of Durham, Durham UK}
\shortauthor{Balogh et~al.}
\shorttitle{Galaxy Evolution}

\suppressfulladdresses
\fulladdresses{
\item M.L. Balogh and R.G. Bower: Department of Physics,
University of Durham, South Road, Durham DH1 3LE UK (m.l.balogh,r.g.bower).
}

\listofauthors{M.L. Balogh, R.G. Bower}
\indexauthor{Balogh, M. L.}
\indexauthor{Bower, R. G.}
\abstract{
The globally-averaged star formation rate in the Universe has been steadily declining
since at least $z\sim 1$.  This may be due either to very local processes operating
within the average galaxy, or to external, environmental effects.  Specifically,
the build-up of structure may be responsible for terminating star formation in some
galaxies and thus decreasing the global average.  We summarize our previous and ongoing
work to  distinguish between these possibilities, by determining the average star formation
rate as a function of redshift and environment, out to $z=0.5$.
}
\addkeyword{galaxies: clusters}

\begin{document}
\maketitle

\section{Introduction}\label{sec-intro}
There is good observational evidence that the total amount of star formation
in the universe has declined substantially over the past few Gigayears \cite{L96,Wilson+02}.
This may be a reflection of local physics on galactic scales, whereby galaxies consume their
gas supply as time progresses, and star formation gradually declines.  However, 
observations show that star formation is inhibited in 
dense environments \cite{B+97,B+98}.  In hierarchical models of galaxy formation, the abundance
of dense clusters increases with time; therefore, perhaps the growth of structure is
partly driving the decline in global star formation.  However, this scenario is
only viable if a suppression of star formation is observed in environments less extreme
than rich clusters, since the latter are too rare to have a significant impact on the
globally averaged star formation rate (SFR).

We can attempt to distinguish between these two interpretations by tracing the SFR
as a function of environment at a series of redshifts.  If the SFR--local density
correlation is independent of redshift, there will be evidence that the global decline
is due to environmental effects, coupled with the hierarchical growth of structure.  
To address this, we have begun a large programme to measure SFRs in different
environments out to $z\sim 0.5$.  The focus is on relatively low-density environments,
since these have not been studied in much detail, and are common enough to contribute significantly 
to the global average.

\section{The Local Universe: SDSS and 2dFGRS}
Recently, data from the SDSS and 2dFGRS 
have allowed the precise measurement of the local SFR-environment correlation.  Both
\scite{2dF-sfr} and \scite{Sloan_sfr} measure a large decrease in the number of strong H$\alpha$-emitting
galaxies in dense regions.  Furthermore, they show evidence that this trend may
be independent of the morphology-density relation \cite{Dressler}.  Both studies
identify a critical density of 1 galaxy ($M_b<-19$) per
Mpc$^2$, below which no further increase in \ha\ strength is observed.  This implies
that environmental effects become important in regions more dense than large galaxy
groups, which contain a substantial fraction of the mass in the Universe.

\section{Intermediate redshifts}
For the purpose of exploring lower mass systems at higher redshifts, we\footnote{In collaboration
with B. Ziegler, R. Davies \& I. Smail.} have 
 selected ten clusters at $z\sim 0.25$ with the lowest detectable X-ray
fluxes, from the catalogue of \scite{V+98}.  For each cluster we have obtained {\it HST} WFPC images
in the F702W filter of the central regions, combined with ground-based
spectroscopy from Calar Alto and WHT over a wider field.  The morphological composition of the
clusters is analysed in \scite{lowlx-morph}, while we consider the spectroscopic
properties in \scite{lowlx-spectra}.  We find that the galaxy populations of these clusters are
remarkably similar to those in more commonly studied clusters, which have masses an order
of magnitude larger (see Figure~1).  This suggests that processes like ram-pressure stripping,
which are only expected to operate in the dense cores of massive clusters, are not
responsible for the environmental dependence of morphology and SFR\@.

It is becoming increasingly clear that direct evidence for environmental effects on
galaxies is likely to be found in dense groups, especially at higher redshift \cite{Kodama_cl0939}.
Therefore, we\footnote{In collaboration with J. Mulchaey, A. Oemler, R. Carlberg and S. Morris.} 
have also begun an ambitious observational
programme using LDSS2 on the Baade (Magellan I) telescope to study galaxy groups at $z\sim 0.45$, selected
from the CNOC2 redshift survey \cite{CNOC_groups}.  The original group catalogues are derived
from a sparse-sampled spectroscopic survey; our programme is designed to complete the spectroscopy
in the fields of these groups, and to probe one magnitude fainter.  To date, a total of 728
spectra in the fields of $\sim 30$ groups have been obtained, in addition to the existing spectroscopy.
From these spectra, we will obtain a complete census of each group, and measure the equivalent 
width of [\ion{O}{2}]$\lambda$3727.  Preliminary
results show that emission lines are much more common amongst group members than in clusters
at similar redshifts.   The mean EW([\ion{O}{2}]) of a preliminary sample of 12 groups is 10.2 \AA,
similar to that of the field at $z\sim 0.3$, but somewhat lower than expected at the mean group
redshift of z=0.45 (see Figure 1).
\begin{figure}[!t]
  \includegraphics[height=0.76\columnwidth]{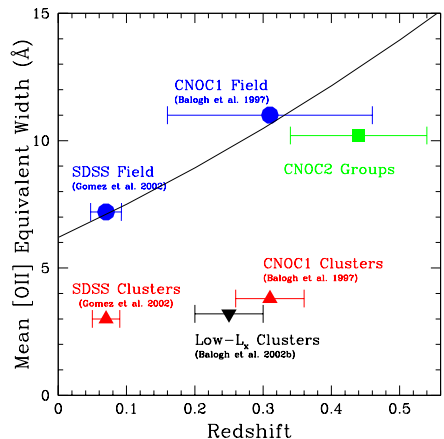}
  \caption{The mean [\ion{O}{2}] equivalent width in each of the samples discussed here.  The results
for the $z\sim 0.4$ groups are based on a preliminary sample of 12 groups.  The other values are based
on published results.  Horizontal error bars show the redshift range of each sample.  Vertical
error bars are omitted, as the statistical errors on the mean are much smaller than the sample
variance (which is large and would clutter the plot).
The solid line shows a scaling of $(1+z)^2$, which is approximately the observed
global rate of evolution \cite{Wilson+02}.}
  \label{fig-envt}
\end{figure}
\section{Summary of Results} 
Our goal is to construct the star formation history of the universe as a function of galaxy
environment. 
Figure~\ref{fig-envt} shows how the mean EW([\ion{O}{2}]) within the virial radius depends on environment
and redshift, for our present sample of groups and clusters. Surprisingly, the amount of
emission in clusters is approximately constant with redshift, so the difference between the cluster and
field SFR {\it increases} with redshift.  This suggests that, at least at cluster densities,
the average SFR is determined by local environment, and not an internal galaxy clock.  On the other hand,
galaxy groups at $z\sim 0.4$ have SFRs only slightly lower than the expected global average at that redshift.  
If SFR is entirely environment-dependent, this means that
the local analogues of these groups should have substantially higher SFRs than the average.

\end{document}